\documentstyle[preprint,revtex]{aps}
\def\gtap{\raisebox{-.4ex}{\rlap{$\sim$}} \raisebox{.4ex}{$>$}}
\def\ltap{\raisebox{-.4ex}{\rlap{$\sim$}} \raisebox{.4ex}{$<$}}
\begin{document}
\thispagestyle{empty}
\font\fortssbx=cmssbx10 scaled \magstep2
\hbox{ %$\vcenter{\special{insert $disk1:[pheno.tex.inputs]uwlogo.imp}}$
%\hskip.2in $\vcenter{
\fortssbx University of Wisconsin Madison} %$ }
\vspace{.3in}
\hfill\vbox{\hbox{\bf MAD/PH/755}
            \hbox{\bf RAL-93-024}
	    \hbox{April 1993}}\par
\vspace{.2in}
\begin{title}
Phenomenological Implications of the $m_t$ RGE Fixed Point \\
for SUSY Higgs
Boson Searches
\end{title}
\author{V.~Barger,$^*$  M.~S.~Berger,$^*$  P.~Ohmann,$^*$ and
R.~J.~N.~Phillips$^\dagger$}
\begin{instit}
$^*$Physics Department, University of Wisconsin, Madison, WI 53706, USA\\
$^{\dagger}$Rutherford Appleton Laboratory, Chilton, Didcot, Oxon OX11  0QX,
England
\end{instit}

\begin{abstract}
\baselineskip=20pt %% doublespace abstract
\nonum\section{abstract}
In minimal SUSY-GUT models with $M_{SUSY}\ltap 1$ TeV, the renormalization
group equations have a solution dominated by the infrared fixed point of the
top Yukawa coupling. This fixed point predicts
$m_t=(200\; {\rm GeV})\sin \beta $;
combined with the LEP results it excludes $m_t\ltap 130$ GeV. For $m_t$
in the range 130-160 GeV, we discuss the
sensitivity of the $m_t$ fixed point result to GUT threshold
corrections and point out the implications for Higgs boson searches.
The lightest scalar $h$ has mass 60-85 GeV and will be detectable at LEPII.
At SSC/LHC, each of the five scalars $h$, $H$, $A$, $H^{\pm }$ may be
detectable, but not all of them together; in one parameter region none will
be detectable.
\end{abstract}

\newpage
For a large top quark mass $m_t>M_W$, the corresponding Yukawa coupling
$\lambda _t$ is expected to be large at the unification scale $M_G^{}$ in
grand unified theories (GUTs). Then the renormalization group equations (RGEs)
cause $\lambda _t$ to evolve rapidly toward an infrared fixed
point at low mass scales\cite{pendleton}.
The prediction for $m_t$ depends on the particle
content below $M_G^{}$. Recent success in achieving gauge coupling
unification based on a low-energy supersymmetry (SUSY)\cite{amaldi,ellis},
with minimal SUSY particle content at $M_{SUSY}^{}\ltap 1$ TeV and
${\rm SU(3)} \times {\rm SU(2)} \times {\rm U(1)}$ evolution below $M_G^{}$,
has stimulated renewed interest in Yukawa coupling unification and fixed
points\cite{ramond,dhr,ross,bbo,knowles,pokorski}. In the present Letter, we
discuss the origins and uniqueness of $m_t$-fixed-point solutions, for the
case that $130<m_t<160$ GeV, and examine the implications for Higgs boson
phenomenology.

{}From the one-loop SUSY standard model RGE
\begin{equation}
{{d\lambda _t}\over {dt}}={{\lambda _t}\over {16\pi ^2}}\left [
-\sum c_ig_i^2+6\lambda _t^2+\lambda _b^2\right ]\;,\label{lambdat}
\end{equation}
with $c_1=13/15$, $c_2=3$, $c_3=16/3$, the couplings evolve toward a fixed
point close to
where the quantity in square brackets in Eq.~(\ref{lambdat}) vanishes.
Here $\lambda _t$ is related to the running mass by
$m_t(m_t)=\lambda _t(m_t)v\sin \beta/\sqrt{2}$ where $\tan \beta =v_2/v_1$ is
the ratio of the two scalar vevs, with $v_1^2+v_2^2=v^2=({\rm 246 GeV})^2$.
If the $\lambda _b$ contribution can be neglected and only the
dominant $g_3$ coupling is retained,
the approximate fixed-point prediction is
\begin{eqnarray}
\lambda _t(m_t)&=&{4\over 3}\sqrt {2\pi \alpha _s(m_t)}
\approx 1\;. \label{mt0}
\end{eqnarray}
giving
\begin{equation}
m_t(m_t)
\approx {v\over \sqrt {2}}\sin \beta \;. \label{mt1}
\end{equation}
More precise two-loop RGE evaluations\cite{bbo} give
\begin{eqnarray}
m_t(m_t)&=&({\rm 192\; GeV})\sin \beta \;, \label{mt2}
\end{eqnarray}
in the regime where $\lambda _b<<\lambda _t$, taking $\alpha _s(M_Z)=0.118$.
(The current experimental average is
$\alpha _s(M_Z^{})=0.118\pm 0.007$\cite{Bethke}.)
The pole mass is related to the running mass by\cite{leut}
\begin{eqnarray}
m_t({\rm pole})&=&m_t(m_t)\left [1+{4\over {3\pi }}\alpha _s(m_t)\right ]
\simeq({\rm 200\; GeV})\sin \beta \;.\label{mt3}
\end{eqnarray}
All subsequent results will be expressed in terms of this pole mass.
Thus in the SUSY-GUT fixed-point solution
$m_t$ is naturally large but dependent on the value of
$\beta $. When evolving from the GUT
scale to electroweak energies, the fixed point is approached
more rapidly from above. If the top quark Yukawa coupling is
below the fixed point at the GUT scale, the convergence to the fixed point is
much more gradual, and in that case
strong statements about the relationship between $m_t$
and $\sin \beta $ cannot be made.

Many SUSY-GUT theories explain the observed $m_b/m_{\tau} $ ratio via a
unification constraint $\lambda _b=\lambda _{\tau }$ at the GUT
scale\cite{CEG}. The one-loop
evolution equation for $R_{b/\tau}\equiv \lambda _b/\lambda _{\tau}$ is
\begin{equation}
{{dR_{b/\tau}}\over {dt}}={{R_{b/\tau}}\over {16\pi ^2}}
\left (-\sum d_ig_i^2+\lambda _t^2
+3\lambda _b^2-3\lambda _{\tau}^2\right )\;,
\label{dRdt}
\end{equation}
with $d_1=-4/3$, $d_2=0$, $d_3=16/3$. If the
bottom quark mass is sufficiently small, then a large top quark Yukawa coupling
is required to counteract the evolution from the gauge couplings; neglecting
$\lambda _t$ would
give a value of $m_b$ too high. Taking the Gasser-Leutwyler\cite{leut} value
$m_b(m_b)=4.25\pm 0.10$ GeV gives rise to the fixed point solution. Larger
values of
the strong coupling constant require a larger top quark Yukawa coupling and
hence yield solutions that are more strongly of the fixed point character.
The uncertainties
on $\alpha _s(M_Z)$, $m_b(m_b)$, and the scale of the supersymmetric
threshold $M_{SUSY}^{}$ introduce a correction in the
coefficient in Eq.~(\ref{mt3}) of up to 10\%. GUT scale threshold corrections
will be discussed below.

There are also fixed-point solutions at $\lambda _b\approx 1$; this
avenue leads to very large $\tan \beta \approx 60$.
In Fig.1 we show SUSY-GUT solution
regions in the $(m_t,\; \tan \beta )$ plane, with the boundary condition
$\lambda _b(M_G^{})=\lambda _{\tau}(M_G^{})$, but without yet imposing any
$m_b(m_b)$ constraint; the
upper and lower bands are the
$\lambda _b$ and $\lambda _t$ fixed-point regions, respectively. In each
case, Yukawa couplings $\lambda _i$ that are $>1$ at the GUT scale evolve
to values in tight agreement with the corresponding fixed-point value at the
electroweak scale. There is a region
where these fixed points overlap, and it was noted in Ref.~\cite{knowles} that
the $m_b/m_{\tau }$ ratio can be obtained from $\lambda _t$,
$\lambda _b$, $\lambda _{\tau}$ fixed points
without necessarily imposing the
unification constraint $\lambda _b=\lambda _{\tau}$.
We adopt the perturbative criterion\cite{bbo} that the
two-loop contribution to the evolution of the Yukawa couplings must not
exceed one quarter of the one-loop contribution; this gives perturbative
bounds $\lambda _t<3.3$ and $\lambda _b<3.1$, eliminating the region
$\tan \beta > 65$.

Threshold corrections at the GUT scale can introduce model-dependent
modifications into the unification criterion $\lambda _b=\lambda _{\tau}$,
in the same way that mass splittings in the GUT scale spectra introduce
corrections to gauge coupling unification.
These corrections are expected to
be larger when $\lambda _t(M_G^{})$ is large.
In the following we impose the low energy boundary condition on
$m_b(m_b)$ and consider relaxations of the
$\lambda _b(M_G^{})=\lambda _{\tau }(M_G^{})$ unification.
In Figure 2(a) we show the effects of
threshold corrections for
$\alpha _3(M_Z^{})=0.118$ taking $m_b(m_b)=4.25$ GeV.
Corrections up to 20\% are
displayed for $\lambda _b(M_G^{})<\lambda _{\tau}(M_G^{})$. Large
threshold corrections
are not possible for $\lambda _b(M_G^{})>\lambda _{\tau}(M_G^{})$ since in
this case the top quark coupling is pushed up against the Landau pole.
In any
case, one sees from Fig. 2(a) that for $\alpha _3(M_Z^{})=0.118$,
threshold corrections even as large as 10\% do not destroy the fixed point
solution. Large threshold corrections have a more severe
impact on the fixed point solutions for significantly
smaller values of $\alpha _s(M_Z)$.
In Figure 2(b) the two-loop evolution
of the top quark Yukawa coupling is plotted for different threshold
corrections along with the value obtained from setting $d\lambda _t/dt=0$
in Eq.~(\ref{lambdat}), which gives an approximation good to about 10\%
to the fixed point.

Once $m_t$ is known, the fixed-point relationship of Eq.~(\ref{mt3})
will uniquely determine
$\sin \beta$. Global analyses of all electroweak
data give a mass $m_t=134^{+19+15}_{-24-20}$ GeV\cite{langacker}.
The present Tevatron lower bounds are $m_t>103$ GeV from
the D0 collaboration and $m_t>108$ GeV from the CDF
collaboration\cite{tevatron}, giving $\sin \beta >0.54$ through
Eq.~(\ref{mt3}). However, there exist possible
top candidate events, one
from D0 and two from CDF, that have no obvious alternative theoretical
interpretations. For the D0 event, a maximum-likelihood analysis is consistent
with a mass in the range $130 < m_t < 160$ GeV (10\% - 90\% interval), and the
production rates in both experiments are consistent with a similar range for
$m_t$\cite{tevatron}. We shall also see below that the fixed point relation
Eq.~(\ref{mt3}) plus LEP Higgs searches exclude $m_t\ltap 130$ GeV.
Given this admittedly
circumstantial evidence, we are motivated to consider the implications if
$m_t$ is indeed in
the range $130 - 160$ GeV, and hence from Eq.~(\ref{mt3}) that
\begin{equation}
0.85<\tan \beta <1.35\;. \label{range}
\end{equation}

In general, the SUSY RGE analysis gives two solutions for $\tan \beta $ at
given $m_t$\cite{ellis,bbo} as shown in Fig.2
(except the value
of $\tan \beta $ is not determined for $m_t$ approaching
200 GeV). However, for
$m_t<160$ GeV, the large $\tan \beta $ solution is
disfavored by perturbative criteria, as shown in Fig.2,
 and by models\cite{ramond,dhr,bbo}
that give $|V_{cb}|=\sqrt{\lambda _c/\lambda _t}$. We therefore
select the $\lambda _t$ fixed point solution.
A range of small $\tan \beta $ ($1\ltap \tan \beta \ltap 5$) is also found
in the standard SU(5) supergravity solutions of Ref.~\cite{lopez,an}, with the
upper bound obtained from proton decay limits; however other
authors\cite{hisano} find a weaker requirement $\tan \beta \ltap 85$ from
considerations of proton decay bounds.

The Higgs spectrum in minimal SUSY models\cite{hhg} consists of two CP-even
scalars $h$ and $H$ ($m_h<m_H$), a CP-odd state $A$ and two charged
scalars $H^{\pm}$; at tree level they are fully described by two
parameters $m_A$ and $\tan \beta$. Important one-loop corrections depend
on $m_t$ and various SUSY parameters (mainly the top-squark mass
$m_{\tilde t}$), giving
mass corrections of order $\Delta m_h^2 \sim
G_F m_t^4 ln(m_{\tilde t}/m_t)$.  One usually selects
typical values of $m_t$ and $m_{\tilde t}$ and then analyzes how the
Higgs masses, couplings and detectable signals vary across the
$(m_A, \tan \beta)$ plane.  Existing LEPI data\cite{lep} already exclude
some areas of this plane.  Extensive analyses \cite{baer,gunion,kz,bcps}
of future possibilities at LEPII and SSC/LHC agree that almost the whole
parameter space can be explored, but that an inaccessible region remains
(with boundary depending on $m_t$ and $m_{\tilde t}$ ) where none of the
Higgs scalars would be discoverable at either LEPII or SSC/LHC.
On the other hand, possible future
$e^+e^-$ colliders above the LEPII range could complete the coverage,
guaranteeing the discovery of at least one Higgs scalar\cite{nlc}.
It is therefore interesting and important to know about any further
theoretical constraints that could reduce the expected range of
parameters and make it easier to confirm or exclude the minimal SUSY
Higgs scenario.  In the present Letter, we point out that the
fixed-point SUSY-GUT solution imposes a severe constraint through
Eq.~(\ref{range}), and we spell out the consequences for Higgs searches at
LEPI, LEPII and SSC/LHC and future higher energy $e^+e^-$ colliders.

(i) LEPI searches.  In the narrow range of Eq.(\ref{range}), the $e^+e^-
\to Z^* h$ channel gives the dominant signals at LEPI. The cross section
differs from that of the corresponding Standard Model (SM) process
by a factor $\sin^2(\beta - \alpha)\gtap 0.7$, where
$\alpha$ is the $h$-$H$ mixing angle, and the detection efficiency
for $h$ decays is approximately the same as that for $H_{SM}$
(if we neglect the possibility of $h \to \tilde Z_1 \tilde Z_1$ decays
to an invisible lightest neutralino).
Hence, with appropriate one-loop
radiative corrections, the combined LEP bound $m(H_{SM}) > 61.0$
GeV \cite{lepsm} can be translated into a bound in the
($m_A, \tan \beta$) plane, or alternatively a bound in the
($m_h, \tan \beta$) plane, making use of Eq.~(\ref{mt3}).
These bounds are shown in Fig.3; they imply that $m_t\gtap 130$ GeV and
$m_h\ltap 85$ GeV.
The left-hand limit on
$m_h$ comes from LEP data; the right-hand $m_h$ limit is the intrinsic
upper limit at given $\tan \beta$.   As higher statistics accumulate,
the reach of LEPI will increase; then either $h$ will be discovered
or the left-hand limit will move to the right.

The bounds in Fig.3 depend somewhat on the one-loop radiative
correction parameters.  We have determined $m_t$ via Eq.~(\ref{mt3})
and have chosen $m_{ \tilde t}=1$ TeV for illustration
(with other SUSY parameters set
as in Ref.\cite{bcps} and playing little role). Lowering
$m_{\tilde t}$ makes the bounds more stringent;
the bound in Fig.3(a) moves right, the left-hand bound in Fib.3(b) becomes
more vertical, and the
right-hand bound in Fig.3(b) moves left. Had we chosen $m_{\tilde t}$
= 0.3 TeV instead, the LEPI bounds would have essentially excluded the entire
range $\tan \beta < 1.35$ and $m_t<160$ GeV of our
fixed-point solutions.

(ii) LEPII searches.  If $m_t\; \ltap \; 160$ GeV,
Figure 3(b) shows that
$60\; {\rm GeV} \ltap m_h \ltap 85$ GeV. Then $h$ will be
discoverable at LEPII\cite{kr}; furthermore, the experimentally difficult
situation $m_h \simeq M_Z$ is unlikely to occur.  In the allowed kinematic
range
of ($m_A,\tan \beta$), we find that $m_{H \pm} \gtap \; 105$ GeV; hence
$H^{\pm}$ will not be discoverable at LEPII.
In principle $A$ could be
discovered via $e^+ e^- \to A h$, but the cross section is too small
through almost all the allowed range, except for a small corner
around $\tan \beta \simeq 1.3$, $m_A \simeq 75$ GeV, $m_h \simeq 60$ GeV.
$H$ production is forbidden by kinematics.

(iii) SSC/LHC searches.  In the allowed region of Fig.3(b), previous
analyses have shown that several Higgs signals will be
viable\cite{baer,gunion,kz,bcps}. Figure
4 shows the signal regions at the SSC from Ref.\cite{bcps} for $h \to \gamma
\gamma$, $H \to 4 \ell$, $A \to \gamma \gamma$ and $H^{\pm} \to \tau \nu$;
a signal from $A \to Zh \to \ell \ell \tau \tau$ will also be
detectable\cite{baer2} in a region somewhat smaller than that for
$A \to \gamma \gamma$.  There appear to be good prospects for detecting
at least one more Higgs scalar $H$ or $A$ or $H^{\pm}$ in addition to the
$h$ detectable at LEPII (the LEPII reach for $h$ is approximated by the contour
$m_h=90$ GeV). Note however that the $h\to \gamma \gamma $ signal
is not expected to be viable for $m_h \ltap 80$ GeV, because of steeply rising
backgrounds\cite{gem}; the lower boundary of this signal region is essentially
the $m_h=80$ GeV contour. Note also that the $H\to 4l$ and
$A\to \gamma \gamma $ (and $A\to Zh$) signals are cut off for
$m_H \simeq m_A \gtap 2m_t$ by competition from $H, A\to t\overline{t}$ decays.
Hence there is a parameter region ($m_A \gtap 2m_t, m_h \ltap 80$ GeV,
$\tan \beta \ltap 1.2$) where no Higgs signal will be detectable at SSC/LHC
and Higgs discovery ($h$ alone) relies on LEPII.

(iv) Higher-energy $e^+e^-$ linear collider searches.  The $m_t$-fixed-point
solutions above have small $\cos ^2(\beta-\alpha) < 0.3$, $0.05$ for
$m_t<160$,
$145$ GeV, respectively, tending toward zero as $m_A$ increases; this factor
suppresses the $e^+e^- \to Ah$, $ZH$ and virtual $WW$, $ZZ \to H$
production channels.
But the channels $e^+e^- \to Zh$, $AH$ have factors $\sin ^2(\beta -\alpha)$
 and are
unsuppressed, while $e^+e^- \to H^+H^-$ has no such factors; copious $h$
production is therefore guaranteed, with $H$, $A$, $H^{\pm }$
too if they are not too heavy.

\acknowledgements
This research was supported
in part by the University of Wisconsin Research Committee with funds granted by
the Wisconsin Alumni Research Foundation, in part by the U.S.~Department of
Energy under contract no.~DE-AC02-76ER00881, and in part by the Texas National
Laboratory Research Commission under grant no.~RGFY9273. PO was supported in
part by an NSF graduate fellowship.

\newpage

{\center \Large Figure Captions}
\vskip 0.5in
%\section*{figure captions}
{\parindent=0in \parskip= 2ex
Fig.1: Contours of constant Yukawa couplings $\lambda _i^G=\lambda _i(M_G^{})$
at the GUT scale in the ($m_t^{\rm pole}, \tan \beta$) plane,
obtained from solutions to the RGE with
$\lambda _t^G=\lambda _b^G$ unification imposed.
The GUT scale Yukawa coupling contours are close together for
large $\lambda ^G$. The
fixed points describe the values of the Yukawa couplings at
the electroweak scale for $\lambda _t^G\gtap 1$ and
$\lambda _b^G\gtap 1$.

Fig.2: RGE results for
$\alpha _s(M_Z^{})=0.118$ with the boundary condition $m_b(m_b)=4.25$
GeV.
\newline
(a) GUT threshold corrections to Yukawa coupling unification.
The solutions strongly exhibit
a fixed point nature, for threshold
corrections $\ltap 10\%$.
Taking a larger supersymmetric threshold $M_{SUSY}^{}$
or increasing $\alpha _s(M_Z)$
moves the curves to the right, so that the fixed point condition becomes
stronger.
(b) Evolution of the top quark Yukawa coupling for
$\tan \beta =1$. The dashed line indicates
${{d\lambda _t}\over {dt}}=0$ which gives an
approximation to the electroweak scale value of $m_t$ with accuracy of
order 10\%.

Fig.3: Fixed-point solution regions allowed by the LEPI data: (a)
   in the $(m_A, \tan \beta )$ plane, (b) in the $(m_h, \tan \beta )$
   plane. The top quark masses are $m_t({\rm pole})$, correlated to
   $\tan \beta $ by Eq.~(\ref{mt3}).

Fig.4: SSC/LHC signal detectability regions, compared with the
   LEPI-allowed region of fixed-point solutions from Fig.3(a) and the
   probable reach of LEPII.
   The top quark masses are $m_t({\rm pole})$.
}

\end{document}